\documentclass[preprint,12pt]{aastex}
             \def \etal {{\it et~al.}}

             \def \s{~\rm{s}}
             \def \km{~\rm{km}}

             \def \AU{~\rm{AU}}

             \def \yr{~\rm{yr}}

             \def \lesssim{\mathrel{<\kern-1.0em\lower0.9ex\hbox{$\sim$}}}
             \def \gtrsim{\mathrel{>\kern-1.0em\lower0.9ex\hbox{$\sim$}}}

\received{2002 December 12}
\begin{document}

%\doublespace

\title{Abundance Anomalies in the X-ray Spectra of the Planetary 
Nebulae NGC 7027 and
BD +30\degr 3639}
\author{Holly L. Maness\altaffilmark{1,2}}
\author{Saeqa D. Vrtilek\altaffilmark{2}}
\author{Joel H. Kastner\altaffilmark{3}}
\author{Noam Soker\altaffilmark{2,4}}

\altaffiltext{1}{Department of Physics, Grinnell College, Grinnell, 
IA 50112; maness@grinnell.edu}
\altaffiltext{2}{HS Center for Astrophysics,  
 Cambridge, MA 02138; svrtilek@cfa.harvard.edu}
\altaffiltext{3}{Center for Imaging Science, 
RIT, 54 Lomb Memorial Drive, Rochester, 
NY 14623; jhk@cis.rit.edu}
\altaffiltext{4}{Department of Physics, Oranim, 
Tivon 36006, Israel; soker@physics.technion.ac.il}

\begin{abstract}

We revisit {\it Chandra} observations of the planetary
nebulae NGC 7027 and BD +30\degr 3639 in order to address
the question of abundance anomalies in the X-ray emitting
gas.   
Enhanced abundances relative to solar of magnesium (Mg) for NGC 7027
and neon (Ne) for BD +30\degr~3639 are required to fit their
X-ray spectra, whereas  
observations at optical and infrared 
wavelengths show depleted Mg and Ne in these systems.  
We attribute the enhancement of Mg in NGC 7027 in the X-ray, 
relative to the optical, to the depletion of Mg onto dust grains
within the optical nebula.
For BD +30\degr~3639, we speculate that the highly enhanced Ne comes
from a WD companion, which accreted a fraction of the wind blown by the
asymptotic giant branch progenitor, and went through a nova-like
outburst which enriched the X-ray emitting gas with Ne.

\end{abstract}

\clearpage

\section{Introduction}

The optically emitting gas of planetary nebulae (PNs) is 
illuminated and ionized by the radiation from the central hot star,
which is evolving to become a white dwarf (WD).
The shaping of the optical nebula, formed from the envelope
of the progenitor asymptotic giant branch (AGB) star,
is a more complicated process, with many open questions
(Kastner, Soker, \& Rappaport 2000).
Of particular interest is whether a binary companion is required
to form non-spherical nebulae, or whether single AGB stars can blow
axisymmetrical winds and form axisymmetrical PNs.
According to Soker \& Rappaport (2000) one of the processes by which a 
companion can shape a PN
is by accreting from the wind of the AGB progenitor; if an accretion
disk is formed around the compact companion, then a jet-like bipolar 
outflow (or collimated fast wind, CFW)
can result.
When the jet material is shocked,
it heats up and may emit in the X-ray band.
Another ingredient in the shaping process is the fast wind from the
central star, blown during the post-AGB and PN stages (Kwok,
Purton, \& Fitzgerald~1978).
The shocked fast wind is expected to emit in the X-ray band.
In particular, an intermediate-velocity  
($\sim 500 \km \s^{-1}$) wind during the post-AGB phase may
explain many of the observed properties of the X-ray emitting gas in
PNs (Soker \& Kastner 2003).

Processes behind the extended X-ray emission are tied to
the shaping processes of PNs (Kastner~\etal~2002), and although not all
shaping processes will lead to X-ray emission, the X-ray
properties of PNs may hint at the nature of their progenitor
(Soker \& Kastner 2003).
With this goal in mind, we present a re-analysis of the
spatially extended X-ray spectra of two PNs, 
BD +30\degr 3639 (PN G064.7+05.5) and
NGC 7027 (PN G084.9-03.4).
Material relevant to these PNs is summarized
in our earlier papers announcing
the detection of extended X-ray emission from them (Kastner~\etal~2000;
Kastner, Vrtilek, \& Soker 2001),
and a study of the X-ray morphologies of these two PNs
(Kastner~\etal~2002).
The observations are described in \S2.  In \S3, we describe our spectral 
analysis, and in \S4
we present the results, emphasizing the large differences
in the abundances of some elements between the X-ray emitting gas and
the optical nebula.
In $\S 4$ we suggest explanations for these findings, focusing
on magnesium (Mg) in NGC 7027, and neon (Ne) in BD +30\degr 3639.

\section{Observations}

{\it Chandra} observed NGC 7027 for 19.0ksec and BD +30\degr 3639
for 18.5ksec with the Advanced CCD Imaging Spectrometer (ACIS)
(Garmire~\etal~1988) as the focal plane instrument.
For both observations the Science Instrument Module
 was translated and the
telescope was pointed such that the telescope 
boresight was positioned
near the center of the spectroscopy CCD array (ACIS
-S); the objects where
imaged on the
central back-illuminated CCD (S3), which provides 
moderate
spectral resolution of $\sim$4.3 at 0.5 keV and 
$\sim$9 at 1.0 keV.
Analyses of these data have already appeared in Kastner
\etal~(2000,2001).
Data from these
observations were reprocessed by the Chandra X-ray
Center (CXC) subsequent to the publication of our
earlier papers and the analyses presented here 
were performed on the reprocessed files. 
Each spectrum was extracted using the 
{\it Chandra}
Interactive Analysis of Observations (CIAO) software 
within a region
judged to contain all the X-ray flux from the 
nebula.  
We note that the current extraction (CIAOv2.2.1) reflects upgrades to the
calibration system that were made after our previous papers
on these systems were published.
The extracted events are 
aspect-corrected, bias-subtracted, graded,
energy-calibrated and  
limited to grade 02346 events (ASCA system).
For both observations, the background count rate 
from a large,
off-source annulus region
(30- and 50-pixel radii) was negligible in 
comparison to the source count rate.

\section{Spectral Analysis}

The low photon statistics and limited
energy range characteristic of these spectra make it difficult to
derive abundances for these objects on the basis of X-ray spectra alone.  
Here we model the spectra using, as guidance, abundances derived from studies in
the optical, ultraviolet, and infrared. 
We used the VMEKAL (Mewe, Lemen, \& van den Oord 1986) code 
as incorporated into XSPEC v11.0
(Arnaud, Borkowski, \& Harrington 1996) to generate model spectra 
that are appropriate for 
optically thin thermal plasma in ionization equilibrium.
When abundance values for elements that are fit by VMEKAL
are not available we set the abundance to Solar.  In this we differ
from our original papers (Kastner~\etal~2000, 2001) where unspecified
abundances were set to zero for elements heavier than Mg. 

\subsection{NGC 7027}

The X-ray spectrum of NGC 7027
shows emission that peaks
at approximately 0.9 keV and 1.3 keV and drops off at approximately
1.4 keV. No significant emission was detected below 0.5 keV and
above 2.0 keV, 
hence we restrict our fits to the 
energy range 0.5-2.0 keV.

Nebular abundances reported by Bernard Salas~\etal~(2001) using ISO and
optical observations produced a fair fit (Fig. 1), 
but do not adequately 
account for features between 0.6-0.7 keV and $\sim$1.3 keV. We attribute 
the feature at 0.65 keV to a blend 
of O lines. We note that while the nebula has been reported to be
somewhat depleted in O 
(Bernard Salas~\etal~2001; Beintema~\etal~1996;
Keyes, Aller, \& Feibelman 1990; Middlemass 1990),
the stellar wind of NGC 7027 is expected to show a greatly enhanced O
abundance (Hasegawa, Volk, \& Kwok 2000). This high O abundance 
could explain the 
feature at 0.6 keV. 
However, we emphasize that, while Kastner \etal~(2001, 2002)
ascribe the X-ray emission to the action of a fast wind, no such
fast wind has yet been detected for this nebula.
The feature 
at $\sim$ 1.3 keV requires enhanced Mg abundance;
we were unable to find alternative explanations for this feature.
We thus fit our X-ray spectrum starting with nebular
abundances as determined by Bernard Salas~\etal~(2001)
 but allowing O and Mg to be free.  The resulting fit (Fig. 2 and
Tables 1,2) 
requires O enhanced to 9 times solar and Mg enhanced to 3 times solar.

In all cases we allowed the intervening column density N$_H$ to be a
free parameter.
Best-fit
values of $N_H$ (Table 2) are consistent with our previous results
and with the typical value of $A_V$ toward NGC 7027
(Kastner \etal~2001,2002);
however, the temperature  
is a factor of $\sim 2.8$ larger, if the abundances of elements
above Mg in NGC 7027 are Solar as assumed here.
A higher temperature implies that the fast wind material that
was shocked and formed the presently X-ray emitting gas was blown at
a somewhat higher speed, and/or the adiabatic cooling, studied
by Soker \& Kastner (2003), was less efficient. "

\subsection{BD +30\degr 3639}

Like NGC 7027, past analysis of BD +30\degr 3639 suggests that this PN's
X-ray emission arises primarily from wind and/or nebular
material (Kastner~\etal~2000). 
Here, we perform several tests 
to determine whether the observed X-ray spectrum is consistent with reported
nebular and/or stellar wind abundances.
The extracted spectrum shows  
distinct peaks near 0.5 keV and 0.95 keV. No significant 
emission was detected below 0.3 keV and above 1.5 keV.
We restrict our fits to the energy range 0.3-1.5 keV.

We first assume nebular abundances reported from optical 
and ultraviolet observations
by Pwa, Pottasch, \& Mo (1986), Aller \& Hyung (1995), and Grevesse \& Sauval
(1998).
The best-fit parameters are listed in Table 2 and
the model is shown in Figure 3. 
We then use abundances reported for the stellar wind
by Arnaud \etal~(1996).
The stellar wind abundances provide a significantly better fit
than the nebular abundances but strong features near 0.95 and 0.6 keV remain
(Figure 4).
We then attempted to fit the X-ray spectrum starting with the stellar wind
abundances but allowing O and Ne to be free parameters.  
This resulted in an acceptable fit ($\chi^2 = 1.5$) with  
O enhanced to 3.6 times solar and
Ne enhanced to 21 times solar.  
Although this Ne enhancement is not predicted by
studies at longer wavelengths, it is consistent with earlier X-ray studies (Arnaud
et al (1996); Kastner et al. (2000))
suggesting that the feature at 0.9 keV can only be fit with an enhanced Ne
abundance.
The model is shown overlaying
the data in Figure 5, and the best-fit parameters are
listed in Tables 1, 2. 
Our best-fit value for T is 0.7 times higher and for 
N$_H$ two times higher than
that obtained by Kastner et al. (2001), whose result for
$N_H$ is consistent with the measured optical obscuration of the
nebula ($A_V$) given a conversion between $N_H$ and $A_V$ appropriate for
the interstellar medium. Our result for $N_H$ would therefore suggest that
the gas to dust ratio in this nebula is elevated over that of the
interstellar medium, along lines of sight toward the X-ray-emitting gas.

\section {Discussion and Conclusions}

NGC 7027 and BD +30\degr3639 show strong
emission features that require enhanced Mg abundance
relative to solar (to fit lines at $\sim$1.3 keV) for NGC 7027,
and enhanced Ne abundance relative to solar (to fit lines at
$\sim$0.9 keV) for BD +30\degr3639
(Kastner~\etal~2000; Kastner~\etal~2001).
Both of these results are surprising since abundance  
values for the nebular and stellar wind material obtained at other
wavelengths indicate depleted Mg and Ne
(Beintama~\etal~1996; Keyes \& Aller 2001; Middlemass 1990).

\subsection {Neon in BD +30\degr3639}

Kastner~\etal~(2000) proposed that the high abundance of Ne found
in BD +30\degr3639 results from the central star which itself is
a carbon-rich Wolf-Rayet ([WC]) object 
(Grosdidier~\etal~2000).
Here we suggest that the source of the large abundance of Ne in
BD +30\degr3639 is a nova eruption on an oxygen-neon-magnesium
(ONeMg) WD companion.
Novae in cataclysmic variables, which are thought to occur on
the surface of ONeMg WDs, are known to eject material with 
Ne abundance up to $\sim 300$ times the solar abundance,
and the total ejected mass is $\gtrsim 10^{-5} -10^{-4} M _\odot$ 
(Starrfield~\etal~1998). The estimated mass in the X-ray emitting 
gas in BD +30\degr3639
is $\sim 2 \times 10^{-5} M_\odot$ (Kastner~\etal~2000). 
Therefore, a nova eruption can easily account for the
$\sim 21$ times solar Ne abundance.
The Mg abundance in such nova eruptions is typically
   a factor of $\sim 5-50$ times lower than the Ne abundance
   (Starrfield et al. 1998).
   Therefore, it is not surprising we do not detect the
   Mg line in the spectrum of BD +30\degr3639. 

The shaping of some bipolar PNs by nova eruption on a WD companion
has been proposed by Soker (2002), who reviews many routes 
for the formation of bipolar PNs, several involving novae eruptions. 
The accretion rate onto the WD cannot be too high in that case, so the
orbital separation cannot be too small.
We do note that an accretion disk may mix Ne in the
ejected jets even when there is steady nuclear burning. 
Observations supporting this proposed scenario
come from symbiotic systems.
Some symbiotic systems are known to experience nova outbursts 
(such systems are referred to as "symbiotic novae";
e.g., RX Puppis, Mikolajewska et al. 1999),
and some bipolar symbiotic nebulae appear to have gone through an
eruptive mass loss event, possibly due to a nova eruption, 
 as in the
cases of V1016 Cyg (Corradi et al. 1999) and He 2-104 (Corradi et al.
2001).
For the symbiotic nova RX Puppis, Mikolajewska et al. (1999) argue for
an orbital separation of $\gtrsim 50 \AU$. 
Although the bright ionized region of BD +30\degr3639
has a general elliptical shape, high resolution
radio molecular-line imaging has revealed the presence of
two, oppositely-directed ``bullets'' of
fast-moving, dense gas, suggestive of the action of collimated
outflows or jets (Bachiller et al. 2000).
To conclude, we propose that the central star of 
BD +30\degr3639 has a massive, $M_{\rm WD} \sim 1 M_\odot$
ONeMg WD companion, at an orbital separation of $\sim 5-50 \AU$. 

\subsection {Magnesium in NGC 7027} 

In principle, peculiar abundance in the X-ray emitting gas 
of NGC 7027 could also result from material ejected from a
WD companion;
indeed, Kastner et al. (2001) attributed the emission peak near 0.9 keV to
Ne lines (whereas here they are modeled as a complex of lines of highly
ionized Fe).
However, we suggest a simpler explanation. 
The nebular abundances derived from previous studies of NGC 7027 are
for the gas in the optical nebulae, and do not include metals locked
in dust particles.
NGC 7027 is known to have a significant amount of dust in the nebular
shell (Sanchez Contreras~\etal~1998) hence it is expected that many metals will
be depleted in the gas phase.
However, in the X-ray emitting gas we don't see this depletion, which
implies that the dust in this medium was either destroyed or was
never formed.

Assuming that the Mg abundance in the X-ray emitting gas is that of
the wind lost by the progenitor, then the depletion in this nebula
is by a factor of $3.0/0.58 \simeq 5$ (Table 1).
Mg depletion in PNs is well established.
For example, in the galactic PN NGC 3918, Harrington, Monk, \& Clegg
(1988) find
depletion by a factor of 3.
Dopita et al. (1997) find Mg depletion by a factor of $\sim 20$ in PNs
in the Magellanic Clouds and note that Mg is not
expected to be formed or destroyed during the evolution of
these PNs (since an ONeMg WD is not formed), and that the magnesium
resides in dust grains which are not destroyed by the 
UV radiation of the central star.
The depletion fraction we find, then, implies that the 
star was born with a Mg abundance of $\sim 3$ times the solar value.
This is feasible, since the distance of NGC 7027 from the galactic
center is similar to that of the sun, and it is close to 
the galactic plane. Mg abundance of $\sim 3$ times solar 
are found in the solar neighborhood 
and closer to the galactic center (e.g., Smartt~\etal~2001).  

In addition, a high Mg abundance due to destruction of dust is
unlikely because the destruction time of the dust in
this PN is likely to be longer than the time elapsed since the star
left the AGB, which we deduce in the following way. 
The destruction time of dust particles in the temperature range
$5 \times 10^5 K \lesssim T_x \lesssim 5 \times 10^7 K$ depends
weakly on the temperature 
(Draine \& Salpeter~1979; Smith et al.~1996). 
However, the destruction time depends on the grain size.
For the PN NGC 3918, for example, Harrington~\etal~
(1988) find that
$\sim$ 50\% of the dust mass is in dust
particles of size $a \gtrsim 0.1 \mu {\rm m}$.
In NGC 7027, the dust grains may be much larger,
with sizes of $a \gtrsim 0.1 \mu {\rm m}$, and up to
$a \sim 5-20 \mu {\rm m}$ (Sanchez Contreras~\etal~1998). 
Scaling from Draine \& Salpeter (1979) for values appropriate here,
in particular grain sizes of $0.1 \mu {\rm m}$ (Jura~1996),
we find the destruction time of dust in NGC 7027 is longer than
$$\tau_d \sim 1500
\left( \frac{n_e}{150 {\rm cm}^{-3}} \right)^{-1}
\left( \frac{a}{0.1 \mu  {\rm m}} \right) \yr$$ 
For an ISM distribution of grain sizes the destruction time at
a pressure of $P\simeq 10^9$ cm$^{-3}$ K, as in the X-ray emitting
gas of NGC 7027, is $\sim 2500$ yr, according to the calculation of
Smith et al.(1996).
The nebular dust cannot be destroyed in the shock, since the shock
moving through the optical-nebular gas is relatively slow.
The dynamical age of NGC 7027 is $<1000$ yr 
(Masson~1989; Latter et al.~2000). 
We conclude that the dust can't be destroyed, even if the nebular
gas is heated to $\sim 10^7$ K.
Because the dust cannot be destroyed for this PN, we speculate that
the X-ray emitting gas results from wind segments which never formed dust.
These segments could result either from a post-AGB wind or a CFW
blown by an accreting companion (for discussion of these possibilities
see Soker 2002).  

\acknowledgements{
We would like to thank our referee, Dr. R. Corradi
for his thoughtful comments.
H.M. was supported by the NSF REU
program at SAO.  S.D.V. was supported in part by NASA Grant NAG5-6711.  
J.H.K. acknowledges support for this research
provided by NASA/CXO grants GO0--1067X and GO2-3009X to RIT. 
N.S. acknowledges support from the US-Israel
Binational Science Foundation and the Israel Science Foundation.}

\section{References}
Aller, L.~H.~\& Hyung, S. 1995, 
\mnras, 276, 1101

Arnaud, K., 
Borkowski, K.~J., \& Harrington, J.~P. 1996, \apjl, 462, L75

Bachiller, R.; Forveille, T.; Huggins, P. J.; Cox, P.; Maillard, J. P.
2000, A\&A 353, L5

Bernard Salas, J., Pottasch, S.~R., Beintema, D.~A., \& Wesselius,
P.~R. 2001, \aap, 367, 949

Beintema, D.~A.~et al. 1996, 
\aap, 315, L253

Corradi, R. L. M., Ferrer, O. E., 
    Schwarz, H. E., Brandi, E., \& Garcia, L. 1999, A\&A, 348, 978

Corradi, R. L. M., Livio, M.,
 Balick, B., Munari, U., \&  Schwarz, H. E. 2001, ApJ, 553, 211 

Dopita, M.~A.~et al. 1997,
\apj, 474, 188

Draine, B.~T.~\& Salpeter, E.~E. 
1979, \apj, 231, 77

Garmire, G.~P., Nousek, J., Burrows, 
D., Ricker, G., Bautz, M., Doty, J., Collins, S., \& Janesick, J. 1988, 
\procspie, 982, 123

Grevasse, N.~\& Sauval, A.~J. 
1998, Space Science Reviews, 85, 161

Grosdidier, Y.,
  Acker, A.i, \& Moffat, A. F. J. 2000, A\&A, 364, 597                  

Harrington, J.~P., 
Monk, D.~J., \& Clegg, R.~E.~S. 1988, \mnras, 231, 577

Hasegawa, T., Volk, K., \& 
Kwok, S. 2000, \apj, 532, 994

Jura, M. 1996, ApJ, 472, 806 

Kastner, J., Li, J., Vrtilek, S., 
Gatley, I., Merrill, K., \& Soker, S. 2002, \apj, in press

Kastner, J., Soker, N., \& Rappaport, S. 2000,
ASP Conference Series,
Vol. 199. Eds. J. H. Kastner, N. Soker, and S. Rappaport

Kastner, J.~H., Soker, N., 
Vrtilek, S.~D., \& Dgani, R. 2000, \apjl, 545, L57

Kastner, J.~H., Vrtilek, 
S.~D., \& Soker, N. 2001, \apjl, 550, L189

Keyes, C.~D., Aller, 
L.~H., \& Feibelman, W.~A. 1990, \pasp, 102, 59

Kwok, S., Purton,
C.~R., \& Fitzgerald, P.~M. 1978, \apjl, 219, L125

Latter, W. B., Dayal, A.,
  Bieging, J. H., Meakin, C., Hora, J. L., Kelly, D. M., \&
  Tielens, A. G. G. M.  2000, ApJ, 539, 783

Masson, C.~R. 1989, \apj, 336, 294

Mewe, R., Lemen, J.R., 
\& van den Oord, G.H.J. 1986, \aaps, 65, 511

Middlemass, D. 1990, \mnras, 244, 294

Mikolajewska, J.,
   Brandi, E., Hack, W., Whitelock, P., A., Barba, R.,  Garcia, L.,
   \& Marang, F. 1999, MNRAS, 305, 190

Pwa, T.~H., Pottasch, S.~R., \& 
Mo, J.~E. 1986, \aap, 164, 184

Sanchez Contreras, C., Alcolea, J., Bujarrabal,
V., \& Neri, R. 1998, \aap, 337, 233

Smartt, S.~J., Venn, K.~A.,
Dufton, P.~L., Lennon, D.~J., Rolleston, W.~R.~J., \& Keenan, F.~P.
2001, \aap, 367, 86

Smith, R.~K., Krzewina, L.~G., Cox, 
D.~P., Edgar, R.~J., \& Miller, W.~W.~I. 1996, \apj, 473, 864

Soker, N. 2002, MNRAS, 330, 481

Soker, N.~\& Kastner, J.~H. 2003, 
\apj, in press, (astro-ph/0209139)

Soker, N.~\&
      Rappaport, S. 2000, ApJ, 538, 241

Starrfield, S., Truran, J.~W., Wiescher, M.~C., \& Sparks, W.~M. 1998, 
\mnras, 296, 502
\clearpage
\normalsize
\centerline{Table 1: Abundances used in fits}
\begin{center}
\begin{tabular}{lccccc}\hline \hline 
&\multicolumn{2}{c}{\bf NGC 7027}&\multicolumn{3}{c}{\bf BD+30$^o$3639}\\ 
Element&BSPBW$^a$
&This paper$^b$&AH$^c$&ABH$^d$ 
&This paper$^e$\\
\hline \hline
He&1.08&1.08&1&1&1\\
C&1.69&1.69&1.1&354&354\\
N&1.71&1.71&1.0&9.1&9.1\\
O&0.55&9.2$\pm2$&0.44&1.26&4.2$\pm0.3$ \\
Ne&0.83&0.83&0.01&10.5&19.3$\pm1.4$\\
Na&0.58&0.58&1&1&1. \\
Mg&0.58&3.0$\pm1$&1 &1&1. \\
Si&0.17&0.17&1&1&1 \\
S&0.51&0.51&0.33&1&1 \\
Ar&0.63&0.63&0.05&1& 1\\
Fe&1&1&1&0&0\\
$\chi_{\nu}^2$&2.0&0.9&9.3&2.1&1.2\\
\hline
\end{tabular}
\end{center}
\noindent
$^a$Bernard Salas \etal~(2001) nebular abundances.

\noindent
$^b$As in BSPBW but with abundances for O and Mg allowed to be free.

\noindent
$^c$Aller \& Hyung (1995) nebular abundances.

\noindent
$^d$Arnaud \etal~(1996) stellar wind abundances.

\noindent
$^e$ As in ABH but with O and Ne allowed to be free.

\vskip 0.5in
\begin{deluxetable}{lcccc}
\tablecolumns{11}
\tablewidth{0pt}
\tablecaption{Model Parameters for Spectral Fits \label{Fits_table}}
\tablehead{\colhead{Object/Model}&\colhead{{\it N$_H$} ( x 10$^{20} 
cm^{-2}$)}&\colhead{{\it T} ( x 10$^6$ K)}&\colhead{Flux (ergs cm$^{-2}$ s$^{-1}$}) 
&\colhead{$\chi_\nu^2$}}
\startdata 
\textbf {NGC 7027} (0.5-2.0 keV) & & & &  \\
Nebular & 70$\pm3$ & 7.9$\pm0.6$ &3.0e-14 & 2.0 \\
Modified Nebular & 41$\pm2$ &8.4$\pm0.6$  &3.6e-14  & 0.9 \\
\\
\textbf {BD +30$^o$ 3639}(0.3-1.5 keV)  & & & & \\
Nebular &74$\pm2$ &1.2$\pm0.02$  & 4.8e-13& 9.3 \\
Stellar Wind& 24$\pm1$ &2.1$\pm0.02$  &6.1e-13  & 2.1 \\
Modified Stellar Wind& 25$\pm1$ & 2.1$\pm0.02$ &6.1e-13  &  1.2 \\
\\
\enddata
\end{deluxetable}

\clearpage

\begin{figure}
\centering
\scalebox{0.5}{\rotatebox{0}{\includegraphics{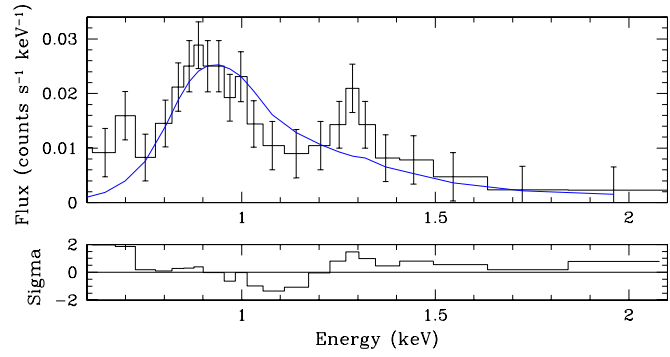}}}
\caption{X-ray spectrum of NGC 7027 with Bernard Salas \etal~(2001) 
nebular model overlaid. 
\label{NGC7027mod_a}}
\end{figure}

\begin{figure}
\centering
\scalebox{0.5}{\rotatebox{0}{\includegraphics{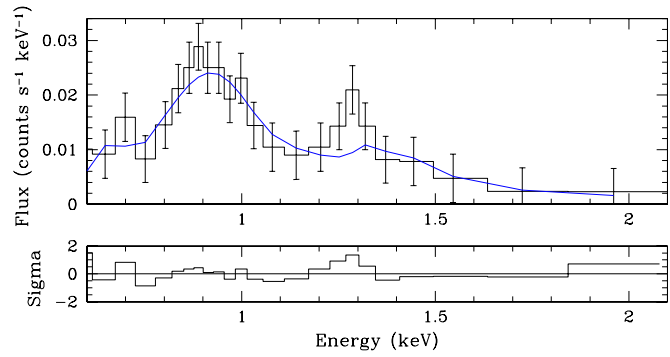}}}
\caption{X-ray spectrum of NGC 7027 with modified nebular model overlaid. 
\label{NGC7027mod_b}}
\end{figure}

\begin{figure}
\centering
\scalebox{0.5}{\rotatebox{0}{\includegraphics{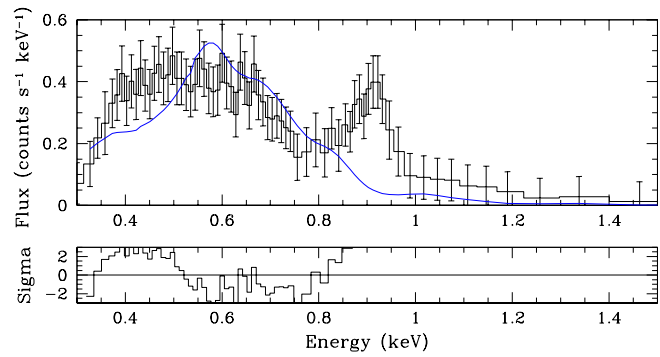}}}
\caption{X-ray spectrum of BD +30$^o$ 3639 with Aller and Hyung (1995) 
nebular model overlaid. 
\label{BD+30b_a} }
\end{figure}

\begin{figure}
\centering
\scalebox{0.5}{\rotatebox{0}{\includegraphics{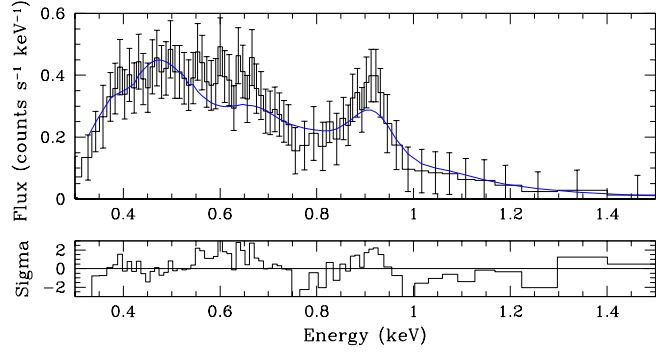}}}
\caption{X-ray spectrum of BD +30$^o$ 3639 with Arnaud \etal~(1996) stellar wind 
model overlaid. \label{BD+30b_b} }
\end{figure}

\begin{figure}
\centering
\scalebox{0.5}{\rotatebox{0}{\includegraphics{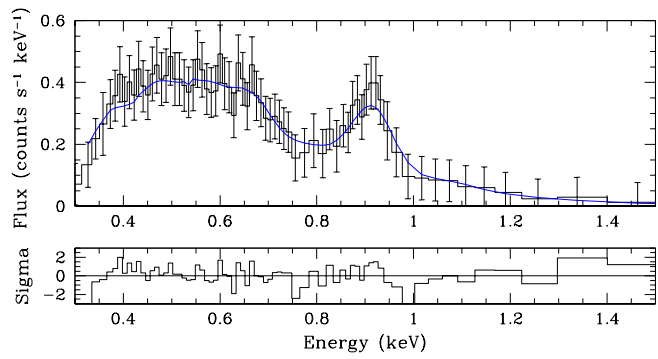}}}
\caption{X-ray spectrum of BD +30$^o$ 3639 with modified stellar wind 
model overlaid. \label{BD+30c} }
\end{figure}

\end{document}